\documentclass[a4paper,fleqn,usenatbib]{mnras}

\usepackage{newtxtext,newtxmath}
\usepackage[T1]{fontenc}
\usepackage{ae,aecompl}

\usepackage{graphicx}	% Including figure files
\usepackage{amsmath}	% Advanced maths commands
\usepackage{amssymb}	% Extra maths symbols

\title[Sensitivity of disk composition]{Protoplanetary disks: Sensitivity of the chemical composition to various model parameters}

\author[V. Wakelam et al.]{
V. Wakelam$^{1}$\thanks{E-mail: valentine.wakelam@u-bordeaux.fr},
E. Chapillon$^{1,2}$,
A. Dutrey$^{1}$, S. Guilloteau$^{1}$, W. Iqbal$^{1}$,
\newauthor
 A. Coutens$^{1}$, L. Majumdar$^{3}$
\\
$^{1}$Laboratoire d'astrophysique de Bordeaux, Univ. Bordeaux, CNRS, B18N, all\'ee Geoffroy Saint-Hilaire, 33615 Pessac, France\\
$^{2}$Institut de Radioastronomie MillimŽtrique (IRAM), 300 rue de la Piscine, 38406 Saint Martin d'H\`eres, France
 \\
$^3$ Jet Propulsion Laboratory, California Institute of Technology, 4800 Oak Grove Drive, Pasadena, CA 91109, USA\\
}

\date{Accepted XXX. Received YYY; in original form ZZZ}

\pubyear{2018}

\begin{document}
\label{firstpage}
\pagerange{\pageref{firstpage}--\pageref{lastpage}}
\maketitle

\begin{abstract}
Protoplanetary disks are challenging objects for astrochemical models due to strong density and temperature gradients and due to the UV photons 2D propagation. In this paper, we have studied the importance of several model parameters on the predicted column densities of observed species. We considered: 1) 2-phase (gas and homogeneous grains) or 3-phase (gas, surface, and bulk of grains) models, 2) several initial compositions, 3) grain growth and dust settling, and 4) several cosmic-ray ionization rates. Our main result is that dust settling is the most crucial parameter. Including this effect renders the computed column densities sensitive to all the other model parameters, except cosmic-ray ionization rate. In fact, we found almost no effect of this parameter for radii larger than 10 au (the minimum radius studied here) except for N$_2$H$^+$. We also compared all our models with all the column densities observed in the protoplanetary disk around DM Tau and were not able to reproduce all the observations despite the studied parameters. N$_2$H$^+$ seems to be the most sensitive species. Its observation in protoplanetary disks at large radius could indicate enough N$_2$ in the gas-phase (inhibited by the 3-phase model, but boosted by the settling) and a low electron abundance (favored by low C and S elemental abundances).
\end{abstract}


%%%%%%%%%%%%%%%%%%%%%%%%%%%%%%%%%%%%%%%%%%%%%%%%%%

%%%%%%%%%%%%%%%%% BODY OF PAPER %%%%%%%%%%%%%%%%%%

\section{Introduction}\label{intro}

Modeling the chemistry in protoplanetary disks is a complex task because of the strong gradients in the physical properties that exist \citep{2014prpl.conf..317D}. Considering the radial and vertical geometries of these objects, the temperature can go from a few 100 K to a few K (excluding of course the most upper atmosphere of the disks where the temperatures are of a few thousand kelvins) while the density varies between $10^4$ and $10^{12}$ cm$^{-3}$. Both the temperature and density radially decrease from the central star. The temperature also decreases towards the mid plane of the disk while the density has an inverse behavior. The chemistry is also driven by the irradiation of UV photons coming from the central star. The consequence of this 2D geometry is a vertical chemical structure. The mid plane is depleted in gas-phase molecules while the outer most layers of the disk are populated by atoms as photodissociation dominates. In between there is a layer of material where we find most of the molecules \citep{2014prpl.conf..317D}. As a consequence, the chemical composition of the gas and ices presents strong variations in the radial and vertical directions.\\
Because of the competition between gas-grain interactions, bimolecular gas-phase chemistry and photodissociation in the molecular layer, the chemical model predictions are very sensitive to the local conditions. There exists a variety of astrochemical models with different types of approximations for the geometry (2D or 1+1D), the chemistry (gas-phase or gas-grain, steady-state or time dependent), the dust properties (single grain size or size distribution), the dynamics (included or not) etc. \citet{2013ChRv..113.9016H} have done a nice compilation of the different existing models. Because of these differences and the high sensitivity of the disk chemistry, it is very difficult to compare one model result to the other.\\
The philosophy of the work presented here is to explore the importance of some of the chemical model approximations and/or parameters on the predicted column densities. We focussed on four different aspects: 
\begin{itemize}
\item[1)] The 2- or 3-phase approximation for surface chemistry: In 3-phase astrochemical models (gas, surface, and bulk of the ice mantles on top of the grains), we make a difference between the first few (usually 2 or 4) most external layers of ices on the grains (considered as the surface) and the rest of the ice below (considered as the bulk). Most of the chemistry occurs at the surface layers while the bulk is protected against evaporation. In the 2-phase models (gas and surface of grains), we do not make any difference between the surface and the bulk. The entire ice is equally chemically active and can desorb. The first question we address is then the impact of these two approximations on the chemical model predictions.
\item[2)] The initial chemical composition: Protoplanetary disks are the result of a long evolutionary sequence starting from the diffuse medium and ending by the formation of stars and planets. During this sequence, the chemistry is never at steady-state. It is very common to simplify the sequence and simply start from an initial chemical composition and use static disk physical properties. The second question we address then is the importance of the initial chemical composition on the disk computed column densities.
\item[3)] The third aspect we explored is the grain growth and dust settling. 
\item[4)] Finally, we tested several values of the cosmic-ray ionization rate based on the idea that this rate may be smaller in protoplanetary disks than the typical value of $10^{-17}$~s$^{-1}$ as claimed by \citet{2015ApJ...799..204C}.
\end{itemize}
This paper is organized as follows. The disk physical and chemical models are presented in section~\ref{models_section} while the tested parameters and approximations are described in section~\ref{tests}. Results are presented in section~\ref{results}, compared with observations in the DM Tau protoplanetary disk in section~\ref{discussion}, and compared with other previous models in section~\ref{comp_models}. Our conclusions are summarized in the last section.

\section{Model description}\label{models_section}

\begin{figure}
\includegraphics[width=1\linewidth]{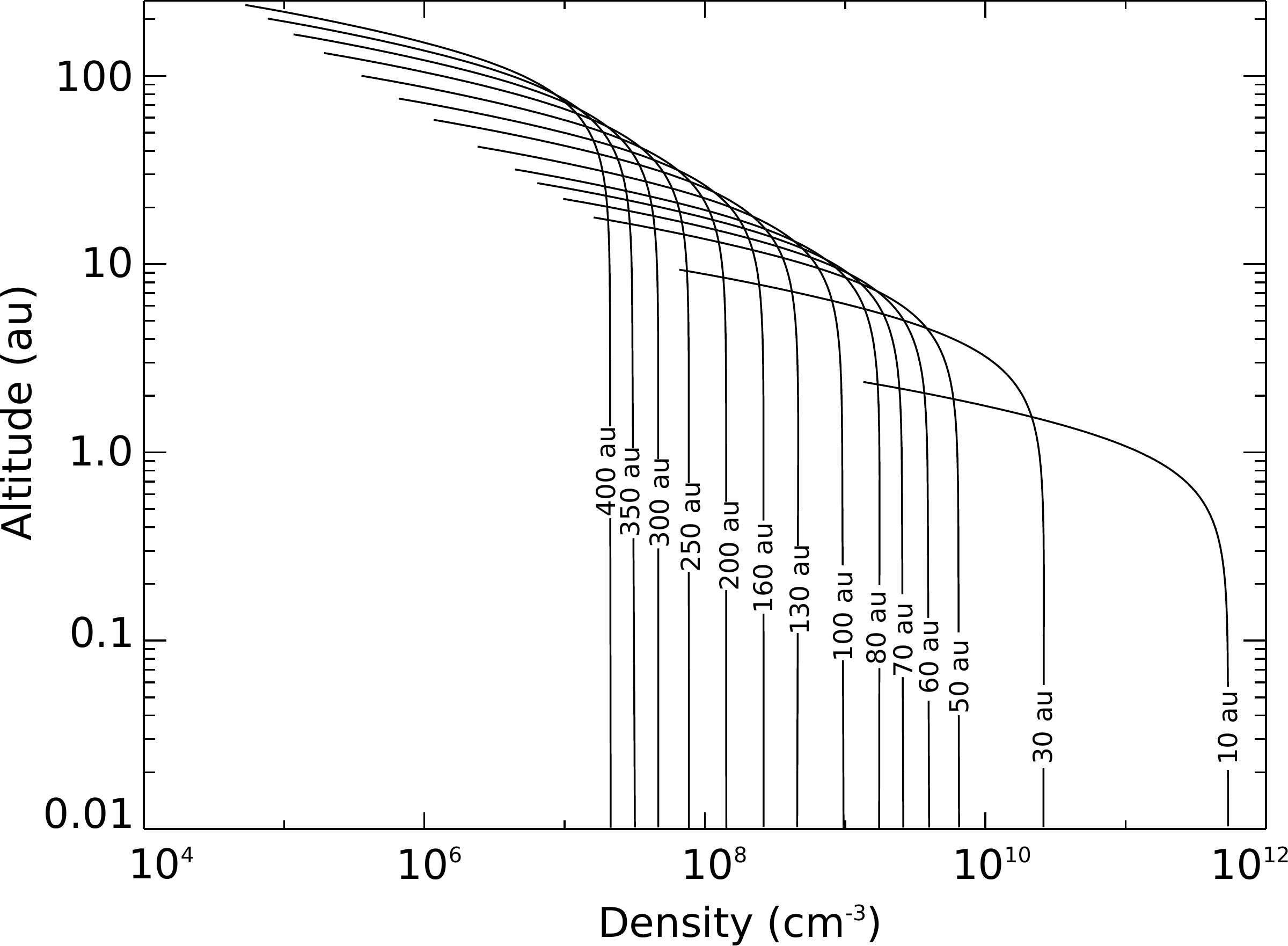}
\caption{Total density vertical profiles at various radius of the disk. One scale height is approximately where the vertical profiles present a bowing.  \label{density}}
\end{figure}

\begin{figure}
\includegraphics[width=1\linewidth]{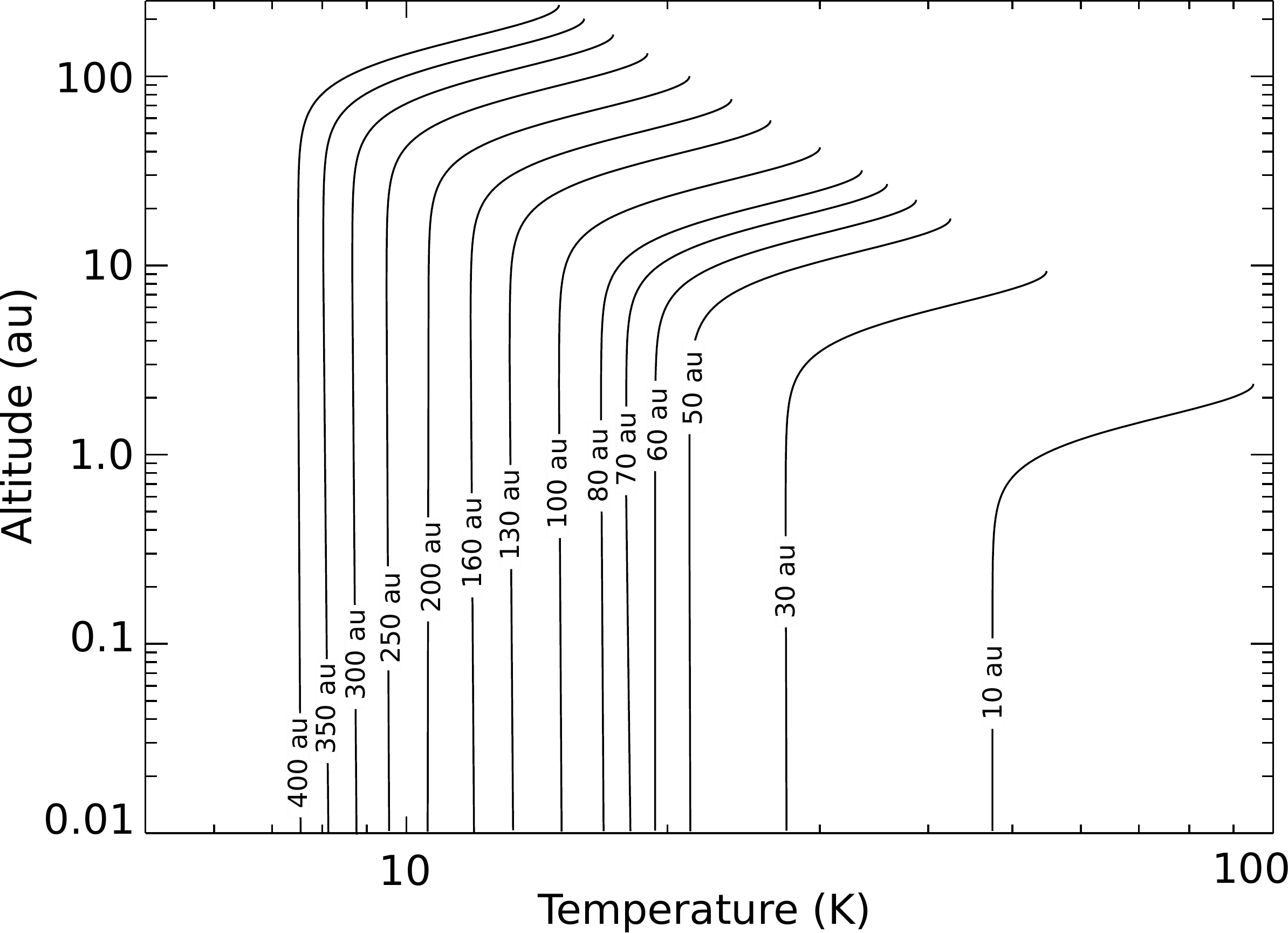}
\caption{Temperature vertical profiles at various radius of the disk. One scale height is approximately where the vertical profiles present a bowing.   \label{temperature}}
\end{figure}

The physical model, we assume, for protoplanetary disks is the same as in \citet{2016A&A...594A..35W}. Briefly, the radial and vertical density profiles are computed assuming that the gas is in hydrostatic equilibrium. The temperature gradients are computed following the prescriptions of \citet{2003A&A...399..773D} and \citet{2014ApJ...788...59W}. For our nominal model (without grain growth), the local visual extinction ($\rm A_v$) is derived assuming  a  conversion  factor of $\rm (A_v/N_H)_0 = 6.25\times 10^{-22}$ \citep{1989MNRAS.237.1019W} with $\rm N_H$ the vertical H column density. The UV flux at a radius $r$ from the star is assumed to be absorbed at the surface of the disk and half of the flux diffuses in the vertical direction. The stellar mass and UV fluxes are similar to those of the DM Tau star and all parameters to compute the 2D physical structure are listed in Table 1 of \citet{2016A&A...594A..35W}.  DM Tau is a TTauri star of mass 0.53 $M_{\odot}$ \citep{2000ApJ...545.1034S}, its Keplerian disk has a total mass of $\sim$0.03 $M_{\odot}$, with a surface density of 0.8 g/cm$^2$ \citep{2011A&A...529A.105G} and an outer radius of 800 au  \citep{1998A&A...339..467G}. \\
The computed density and temperature gradients are shown in Figs.~\ref{density} and  \ref{temperature} respectively. Throughout the paper, $r$ refers to the radius from the star and $z$ is the altitude above mid plane, both are in unit of au. $h$ is the scale height $h = \sqrt{\frac{k_B \times T_{mid} \times r^3}{\mu \times amu \times G \times M_*}}$, with $k_B$ the Boltzmann constant, $T_{mid}$ the mid plane temperature, $\mu$ the mean molecular weight of the gas, $amu$ is the atomic mass unit, G the gravitational constant, and $M_*$ the mass of the star. We assume a duration of $4\times 10^6$~yr for the chemical evolution. Although the chemistry in the disk has not reached steady state (even in the mid plane where the gas-grain exchanges dominates the chemistry), it only slowly evolves  in the disk for ages beyond $10^6$~yr for all the conditions (including the cosmic-ray ionization rates) used in this study. For this reason, the results presented here would still stand if the disk were two times younger or older.

The 2D chemical composition in the disk is computed with the gas-grain code Nautilus fully described in \citet{2016MNRAS.459.3756R} and using the physical structure described above as input parameters. The chemical composition of the gas and the volatile ice on top of interstellar dust grains is computed as a function of time by solving a set of differential equations. The gas-phase chemical network is derived from kida.uva.2014 \citep{2015ApJS..217...20W}, which contains all relevant and common processes for the gas-phase chemistry (unimolecular reactions, such as photo-dissociation and ionisation by direct UV photons and comic-ray induced UV photons, bimolecular reactions such as ion-neutral and neutral-neutral reactions, and electronic recombination). This network has been updated from more recent publications from \citet{2015MNRAS.453L..48W}, \citet{2016MNRAS.456.4101L}, \citet{2016MolAs...3....1H}, \citet{2017MNRAS.469..435V}, and \citet{2017MNRAS.470.4075L}. CO, H$_2$, and N$_2$ self-shielding are computed using tabulated values from \citet{1996A&A...311..690L}, \citet{2009A&A...503..323V}, and \citet{2013A&A...555A..14L} respectively.\\
 In the 3-phase version of the code, the surface layer is defined by the two most external layers of molecules while the rest below defines the bulk of the mantle. Species from the gas-phase can be physisorbed on the grain surfaces upon collision while physisorbed species can desorb through thermal and non-thermal processes. For non-thermal desorption, the model includes the desorption of species due to cosmic-ray induced heating \citep{1993MNRAS.261...83H}, (direct and indirect) UV photons, and exothermic surface reactions (a.k.a. chemical desorption). The surface network is originally the one included in the gas-grain model developed in Prof. Herbst's team with modifications described in \citet{2015MNRAS.447.4004R}, \citet{2015MNRAS.453L..48W}, \citet{2016MNRAS.456.4101L}, \citet{2016MolAs...3....1H}, \citet{2017MNRAS.469..435V}, and \citet{2017MNRAS.470.4075L}. The binding energies have been updated from \citet{2017MolAs...6...22W}. All the data on the surface reactions and binding energies have been included in the online KIDA database \footnote{http://kida.obs.u-bordeaux1.fr/}.  For photo-desorption, considering the complexity of the mechanism, we followed the advice of \citet{2013ApJ...779..120B} and used a constant photo-desorption yield of $10^{-4}$ molecule/photon for all molecules \citep[see also][for discussions]{2016MNRAS.459.3756R}. For the chemical desorption, we are using the model described by \citet{2007A&A...467.1103G} with a $a$ parameter of 0.1 (which implies that nearly 1\% of the singly produced species desorbs).  Both the surface and the bulk are chemically active but the binding and diffusion barrier are set differently so that the reactivity in the bulk is much slower: the diffusion-to-binding energy ratio is 0.4 on the surface and 0.8 in the bulk. In addition, in the bulk, the binding energies, smaller than the one of water, are set to the one of water (except for H,
H$_2$, C, N and O) in agreement with the findings of \citet{2015PCCP...1711455G}. Diffusion of species occurs through thermal hopping except for oxygen, which has been shown to diffuse through tunneling effects \citet{2013PhRvL.111e3201M}. Without this tunneling diffusion and with the new estimate of the high binding energy of O on water surfaces, complex organic molecules would not be produced efficiently on the surfaces. The reaction-diffusion competition as proposed by \citet{2007A&A...469..973C} and \citet{2011ApJ...735...15G} is included in the model. 
The bulk and the surface phases interact in the sense that the species from the surface are incorporated little by little to the bulk while new species land on the surfaces. Similarly, the bulk species become the surface ones while surface species desorb in the gas-phase. All details on how the processes are included in the code and the model parameters are given in \citet{2016MNRAS.459.3756R}.

\section{Tests}\label{tests}

\begin{figure}
\includegraphics[width=1\linewidth]{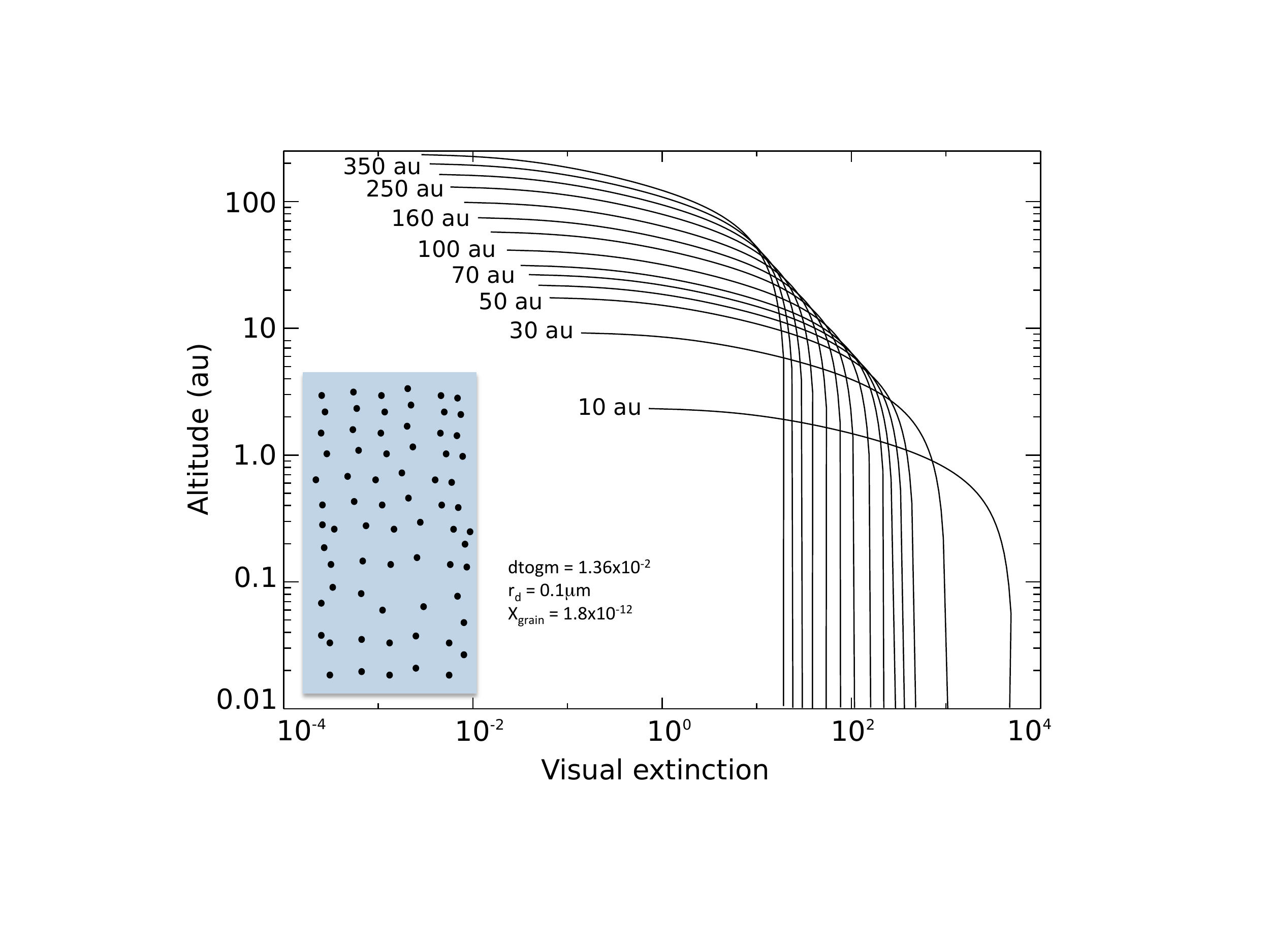}
\includegraphics[width=1\linewidth]{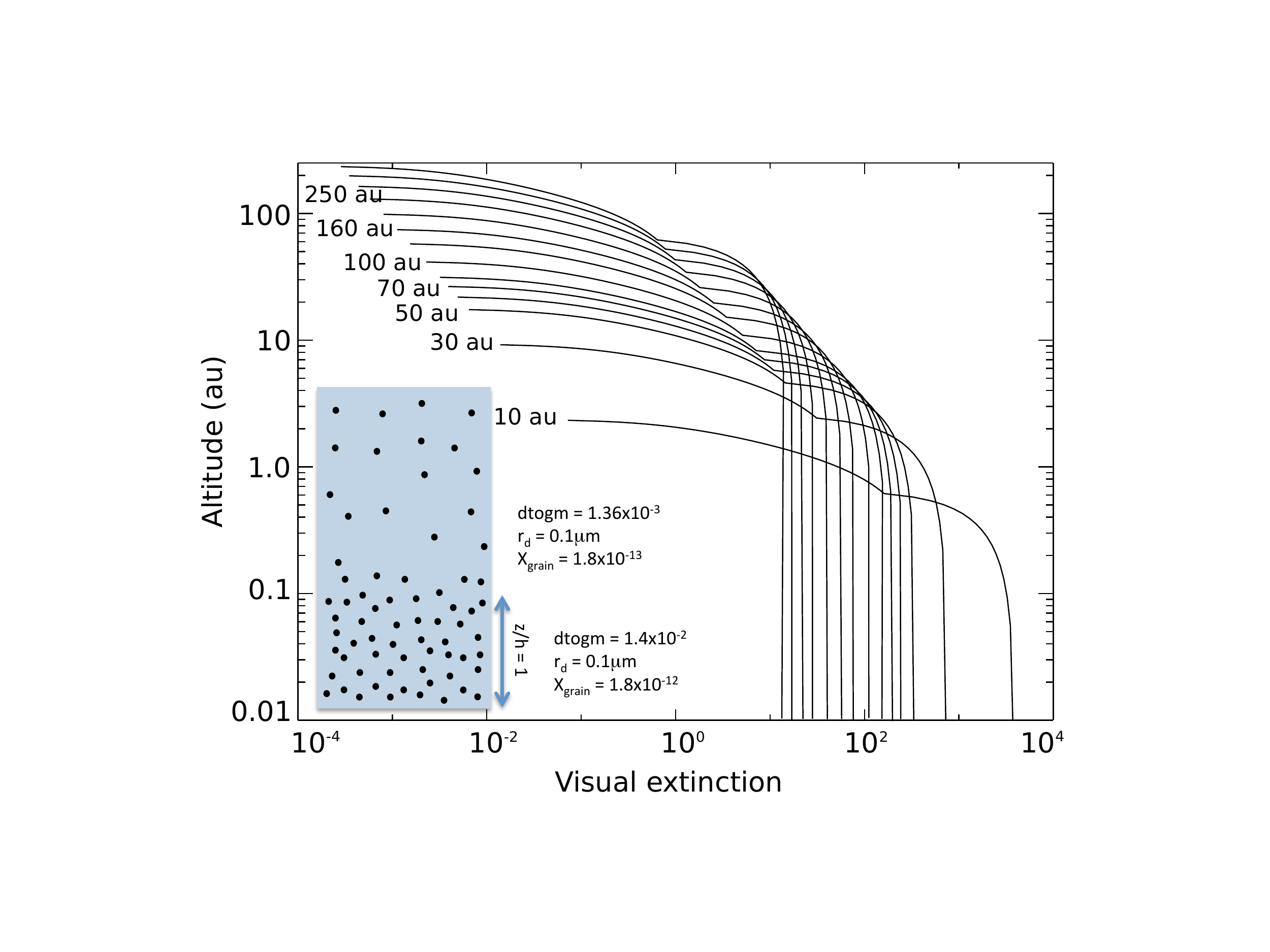}
\includegraphics[width=1\linewidth]{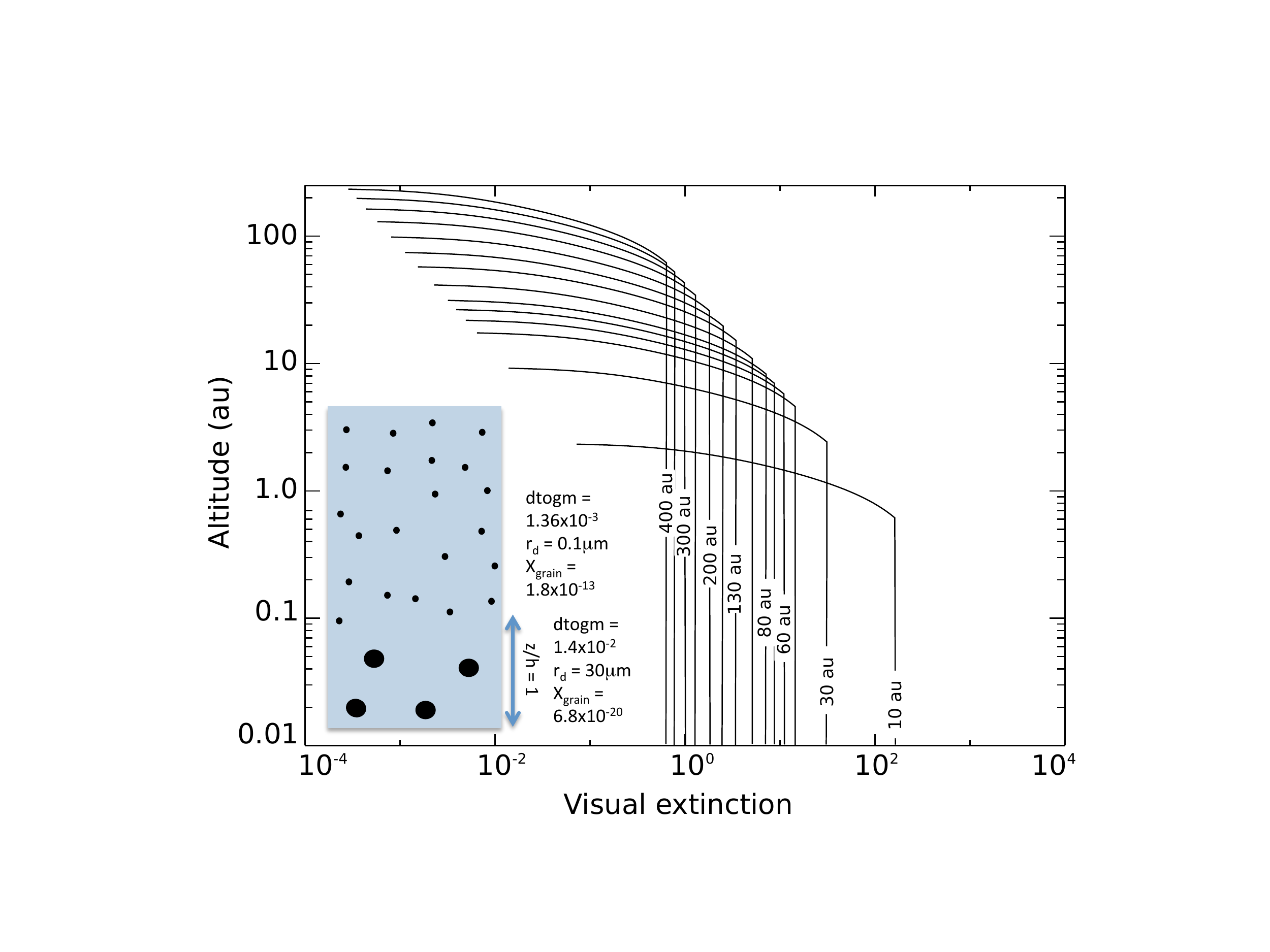}
\caption{Computed visual extinction in the case of homogeneous 0.1$\micron$ size grains (upper panel), settling without grain growth on the mid plane (middle panel), and settling with grain growth on the mid plane (lower panel). One scale height is approximately where the vertical profiles present a bowing in the upper panel while it is when they present a discontinuity in the two other panels. $\rm X_{grain}$ is the abundance of grains with respect to the total proton density.  \label{Av_models}}
\end{figure}

We have tested the effect of several assumptions commonly used for disk chemical modeling. 
The following models have been run:
\begin{itemize}
\item[1)] 2- or 3- phase model: The chemical model is either computed with the 2-phase or 3-phase approximation. Although the 3-phase models are physically more realistic, they are numerically difficult to perform. As a consequence, 2-phase models are still commonly used for protoplanetary disk chemical modeling but also for many other types of applications.
\item[2)] Initial conditions: The disk is usually assumed to form from a cold dense core. The initial chemical composition is usually assumed to be either atomic \citep[for instance][]{1998A&A...338..995W} or molecular \citep[for instance][]{2009A&A...493L..49H,2016ApJ...832..110C}. 
Here we will consider three models. In the first one, all the gas is assumed to be atomic except for H$_2$ \citep[with atomic abundances listed in Table 1 of][except for sulphur assumed to have the undepleted abundance of $1.5\times 10^{-5}$]{2017MNRAS.469..435V}. This is the typical initial condition for a cold core forming from diffuse medium (where hydrogen is usually already converted to H$_2$). In the two other models, we first compute the composition of a cold shielded core  at 10~K, with a H density of $2\times 10^4$~cm$^{-3}$, and a cosmic-ray ionization rate of $10^{-17}$~s$^{-1}$ (the grains remain at 0.1$\micron$ for this stage). The cold core chemical compositions at $10^5$~yr and $10^6$~yr are then used as initial compositions for each of the two models as these times represent the chemical ages based on the comparison between chemical models and observations \citep[see for instance][]{2017MNRAS.470.4075L,2017MNRAS.469..435V}. This simple assumption of course does not take into account the evolution of the chemistry between the cold core phase and the protoplanetary disk itself \citep{2016MNRAS.462..977D}.

\item[3)] Dust settling and grain growth: 
The model is first run assuming a homogeneous distribution of the grains: grains are the same at all altitudes and all radii (grain radius $r_d$ of 0.1$\mu$m and dust-to-gas mass ratio of $10^{-2}$). Models are also run to simulate the dust settling without and with grain growth. Dust settling depends on the product of the angular velocity and the dust stopping timescale, which scales as the particle size. When this product  is larger than 1, particles decouple from the gas. 
Large grains are assumed to settle by being distributed with a smaller scale height than the gas. The settling ratio, ratio of the dust scale height to the gas scale height, is approximated in the same way as by \citet{2013MNRAS.431.1573B}. These approximations are based on the numerical study of \citet{2009A&A...496..597F}. The long wavelength dependence of the mm continuum emission of disks suggests that grains have grown to sizes up to several cm \citep[e.g.][and references therein]{2010A&A...521A..66R}. To evaluate a case of grain growth, we used an effective grain size of 30 $\mu$m, which presents the same cross-section as a size distribution with radii between  0.01 $\mu$m and 10 cm and an exponent of 3.5 for the number density as a function of size, a value found in the ISM \citep{1977ApJ...217..425M}.
 In our model, dust settling is simulated by changing the dust-to-gas mass ratio ($dtogm$), in the upper layer of the disk (above a certain altitude $z$ that we will call $z_{t}$), to $10^{-3}$ and putting the extra mass below $z_{t}$ (the average $dtogm$ is still $10^{-2}$). The grain growth is simulated by replacing grains below $z_{t}$ with larger 30$\mu$m grains. Both the dust settling and the grain growth changes the number density of grains in the vertical direction. The visual extinction is also changed as the  $ A_v/N_H$ conversion factor is scaled as follows: $$(A_v/N_H) = (A_v/N_H)_0 \frac{dtogm}{10^{-2}} \frac{10^{-5}}{r_d (cm)}$$ 
 Since large grains are significantly larger than the optical wavelengths,we assume that the $(A_v/N_H)$ factor scales with the dust size (equation above).
Note that in all cases, we do not include a grain size distribution and for each grid of the model we have only one single size (that can change though). Fig.~\ref{Av_models} shows the visual extinction obtained in the three types of models described here for $z_{t}$ = 1$h$. The computed visual extinction in the three models are the same above $z_{t}$. Below this point, in the case of grain growth, $A_v$ remains constant below this layer as the big grains are not contributing to the visual extinction anymore while in the case without grain growth, the visual extinction is increased because of a large amount of small grains. 

\item[4)] Cosmic-ray ionization rate: \citet{2013ApJ...772....5C,2014ApJ...794..123C,2015ApJ...799..204C} propose observational and theoretical evidence that the cosmic-ray ionization rate in protoplanetary disks may be smaller than the typical interstellar value due to exclusion by the stellar winds. To test the effect of decreasing the cosmic-ray ionization rate, we run the model with two values of $\zeta$: 
$10^{-17}$ and $10^{-19}$~s$^{-1}$ (same value in the entire disk). The lower value is the upper limit suggested by \citet{2015ApJ...799..204C}, which is based on the exclusion of Galactic cosmic-rays by TTauri wind, and in that case the dominant source of ionization in the disk mid plane is the decay of short-lived radionuclides. The upper value is the one commonly used in astrochemical models.
\end{itemize}

The models we have run are summarized in Table~\ref{models}. We did not make a full grid of models because of computational capabilities but selected some of the models as examples. For the settled and grain growth models, for example, we have used only the age of $10^6$~yr for the molecular initial conditions because these conditions are the most evolved ones as compared to the atomic one, and so may produce the largest differences. For the cosmic-ray ionization rate, we also limited our tests to the two models presenting the largest differences.

\begin{table*}
	\centering
	\caption{Summary of the different models.}
	\label{models}
	\begin{tabular}{lccccc} % four columns, alignment for each
		\hline
		\hline
	& Model type & Initial conditions & Settling & Grain growth & $\zeta$ (s$^{-1}$) \\
	\hline
A1 & 2-phase & Atomic & no & no & $10^{-17}$ \\
A2 & 2-phase & Mol. $10^5$~yr & no & no & $10^{-17}$ \\
A3 & 2-phase & Mol. $10^6$~yr & no & no  & $10^{-17}$ \\
\hline
B1 & 3-phase & Atomic & no & no & $10^{-17}$ \\
B2 & 3-phase & Mol. $10^5$~yr & no & no & $10^{-17}$ \\
B3 & 3-phase & Mol. $10^6$~yr & no & no & $10^{-17}$ \\
\hline
C1 & 2-phase & Atomic & yes & yes & $10^{-17}$ \\
C2 & 2-phase & Mol. $10^6$~yr & yes & yes & $10^{-17}$ \\
\hline
D1 & 3-phase & Atomic & yes & yes & $10^{-17}$ \\
D2 & 3-phase & Mol. $10^6$~yr & yes & yes & $10^{-17}$ \\
\hline
E1 & 3-phase & Atomic & yes & no & $10^{-17}$ \\
E2 & 3-phase & Mol. $10^6$~yr & yes & no & $10^{-17}$ \\
\hline
F1 & 2-phase & Atomic & no & no & $10^{-19}$ \\
F2 & 3-phase & Mol. $10^6$~yr & yes & yes & $10^{-19}$ \\
		\hline
	\end{tabular}
\end{table*}

\section{Results}\label{results} 

Considering the large number of models we have run and the large number of molecules we can look at, we focus this paper to the molecules observed in the protoplanetary disk around DM Tau. Table~\ref{table_obs} lists the observed molecules with their observed column densities at 300 au. 

\begin{table*}
	\centering
	\caption{Observed vertical column densities (in cm$^{-2}$) in DM Tau at 300 au.  E2.1 to E2.6 refers to the results of several additional models described in section~\ref{comp_obs}.}
	\label{table_obs}
	\begin{tabular}{llccccccc} % four columns, alignment for each
		\hline
		\hline
		Molecule & Observed value & E2 & E2.1 & E2.2 & E2.3 & E2.4 & E2.5 & E2.6  \\
		\hline
CO & $(1.4 \pm 0.6)\times 10^{17} $ & $2.8\times 10^{17}$ &   $1.4\times 10^{17}$ &    $3.3\times 10^{17}$ &  $2.1\times 10^{18}$ & $2.1\times 10^{18}$ &  $2.1\times 10^{18}$ & $8.9\times 10^{18}$ \\
HCO$^+$ & $(1.1 \pm 0.3)\times 10^{13a}$ &   $6.7\times 10^{10}$ &  $3.1\times 10^{11}$ &  $9.7\times 10^{10}$ & $4.9\times 10^{12}$ &  $5.0\times 10^{12}$ & $3.3\times 10^{12}$ & $3.1\times 10^{10}$ \\
  & $4\times 10^{12b}$ &   &   &   &  &   &  &  \\
N$_2$H$^+$ & $(1.1 \pm 0.3)\times 10^{11} $  &  $6.3\times 10^{8}$ &  $1.2\times 10^{9}$ &   $6.9\times 10^{8}$ &    $ 3.5\times 10^7$ &     $4.4\times 10^7$ &     $3.9\times 10^5$ & $6.2\times 10^7$ \\
CCH & $(2.8 \pm 0.2)\times 10^{13} $ & $4.1\times 10^{13}$ & $5.9\times 10^{13}$ & $3.7\times 10^{13}$ & $4.7\times 10^{14}$ & $4.7\times 10^{14}$ & $9.5\times 10^{14}$ & $3.7\times 10^{13}$\\
HCN & $(6.5 \pm 0.9)\times 10^{12} $ &    $3.1\times 10^{13}$ &  $2.6\times 10^{14}$  &  $1.5\times 10^{14}$ &  $3.9\times 10^{13}$ &  $3.9\times 10^{13}$ &  $1.7\times 10^{12}$ &   $6.5\times 10^{13}$ \\
CN & $(3.5 \pm 0.9)\times 10^{13} $  &    $4.8\times 10^{14}$ &  $8.4\times 10^{14}$ &   $7.5\times 10^{14}$ &   $5.5\times 10^{14}$ &  $5.5\times 10^{14}$  &  $4.6\times 10^{13}$ &  $4.6\times 10^{14}$\\
CS & $(3.5 \pm 0.1)\times 10^{12} $  &   $1.3\times 10^{15}$ &   $1.7\times 10^{15}$ &  $3.6\times 10^{15}$ & $4.5\times 10^{14}$ &  $4.5\times 10^{14}$ &  $3.3\times 10^{13}$ &  $2.2\times 10^{15}$ \\
SO & $\le 7.5\times 10^{11} $   & $5.0\times 10^{12}$ &   $5.0\times 10^{13}$ &  $3.1\times 10^{13}$ & $3.3\times 10^{12}$ &  $3.3\times 10^{12}$ &  $3.5\times 10^{11}$ &  $7.3\times 10^{14}$ \\
H$_2$S & $\le 1.4\times 10^{11} $  &  $5.3\times 10^{14}$ &  $2.5\times 10^{15}$ &  $7.6\times 10^{13}$ &  $5.1\times 10^{13}$ &  $5.1\times 10^{13}$ & $4.7\times 10^{12}$ &  $3.7\times 10^{12}$ \\
CCS & $\le 1.1\times 10^{12} $  &    $2.7\times 10^{13}$ &  $3.5\times 10^{13}$ &  $3.7\times 10^{13}$ & $3.2\times 10^{13}$ & $3.2\times 10^{13}$ & $3.7\times 10^{12}$ & $2.0\times 10^{13}$ \\
HC$_3$N & $\le 7.5\times 10^{11} $ &   $1.0\times 10^{11}$ &  $2.5\times 10^{12}$ &  $4.5\times 10^{11}$ &  $1.9\times 10^{12}$ & $2.0\times 10^{12}$ & $9.1\times 10^{10}$ & $6.9\times 10^{10}$\\
		\hline
	\end{tabular}
	\\
 References for the observed column densities are the following: \citet{2011A&A...535A.104D} for CS, SO and H$_2$S,  \citet{2012ApJ...756...58C} for HC$_3$N and CCS, \citet{2012A&A...537A..60C} for CN and HCN, \citet{2007A&A...464..615D} for N$_2$H$^+$, \citet{2010ApJ...714.1511H} for CCH, \citet{2007A&A...467..163P} for HCO$^{+}$ (value indicated with the a exponent) and CO, and \citet{2015A&A...574A.137T} for HCO$^+$ (value indicated with the b exponent). All the observed surface densities have been derived using a proper radiative transfer disk model \citep[DISKFIT, see][for details]{2007A&A...467..163P,2011A&A...535A.104D}.  
\end{table*}

\subsection{With homogeneous 0.1$\micron$ grains (models A and B)}\label{without_sed}

\begin{figure*}
\includegraphics[width=0.9\linewidth]{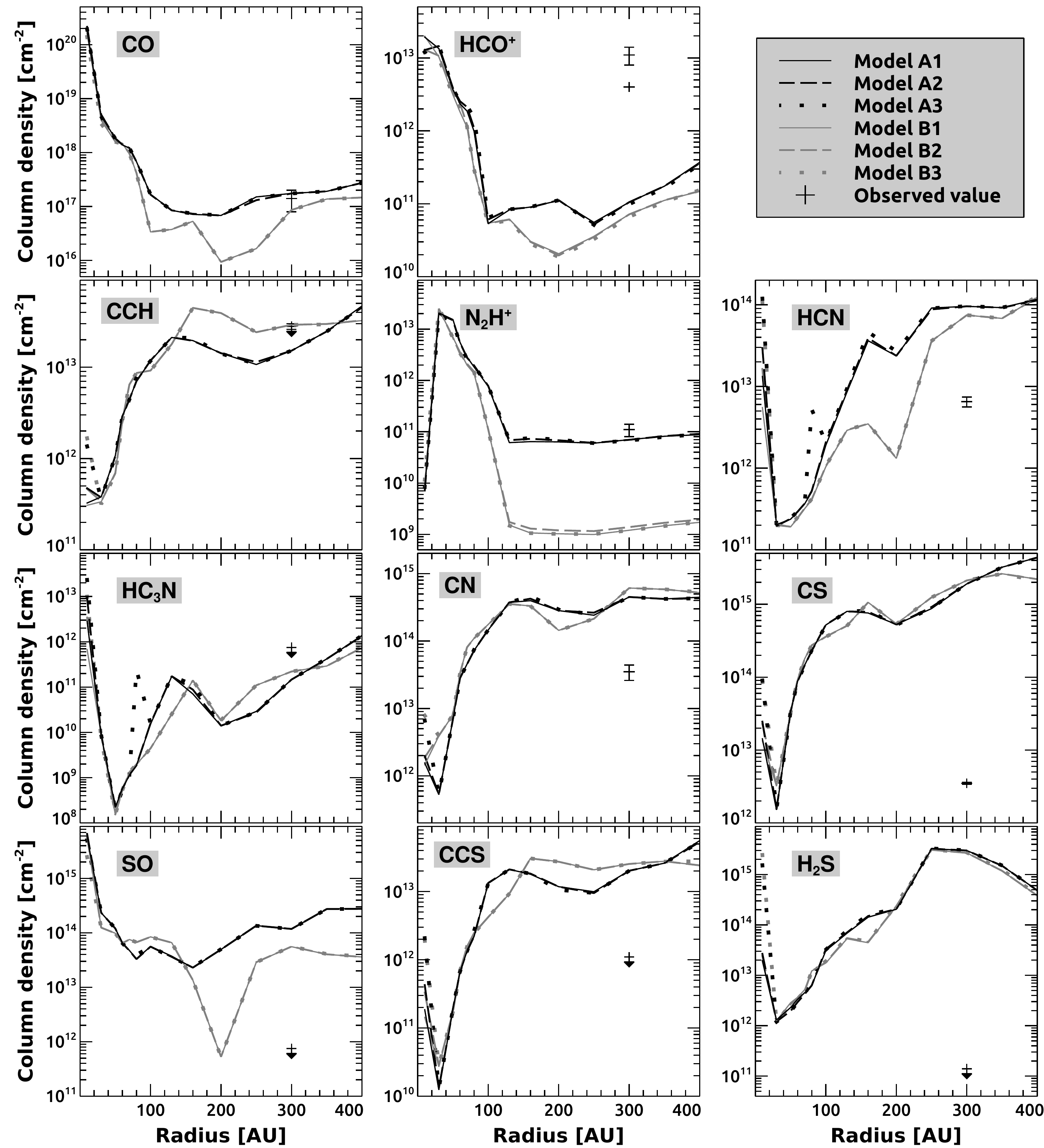}
\caption{Column densities of a selection of gas-phase species as a function of radius to the central star for 2-phase models  A(1 to 3) and 3-phase models B(1 to 3). The points at 300 au represents the observed column densities. \label{allmols_nosed}}
\end{figure*}

\begin{figure}
\includegraphics[width=1\linewidth]{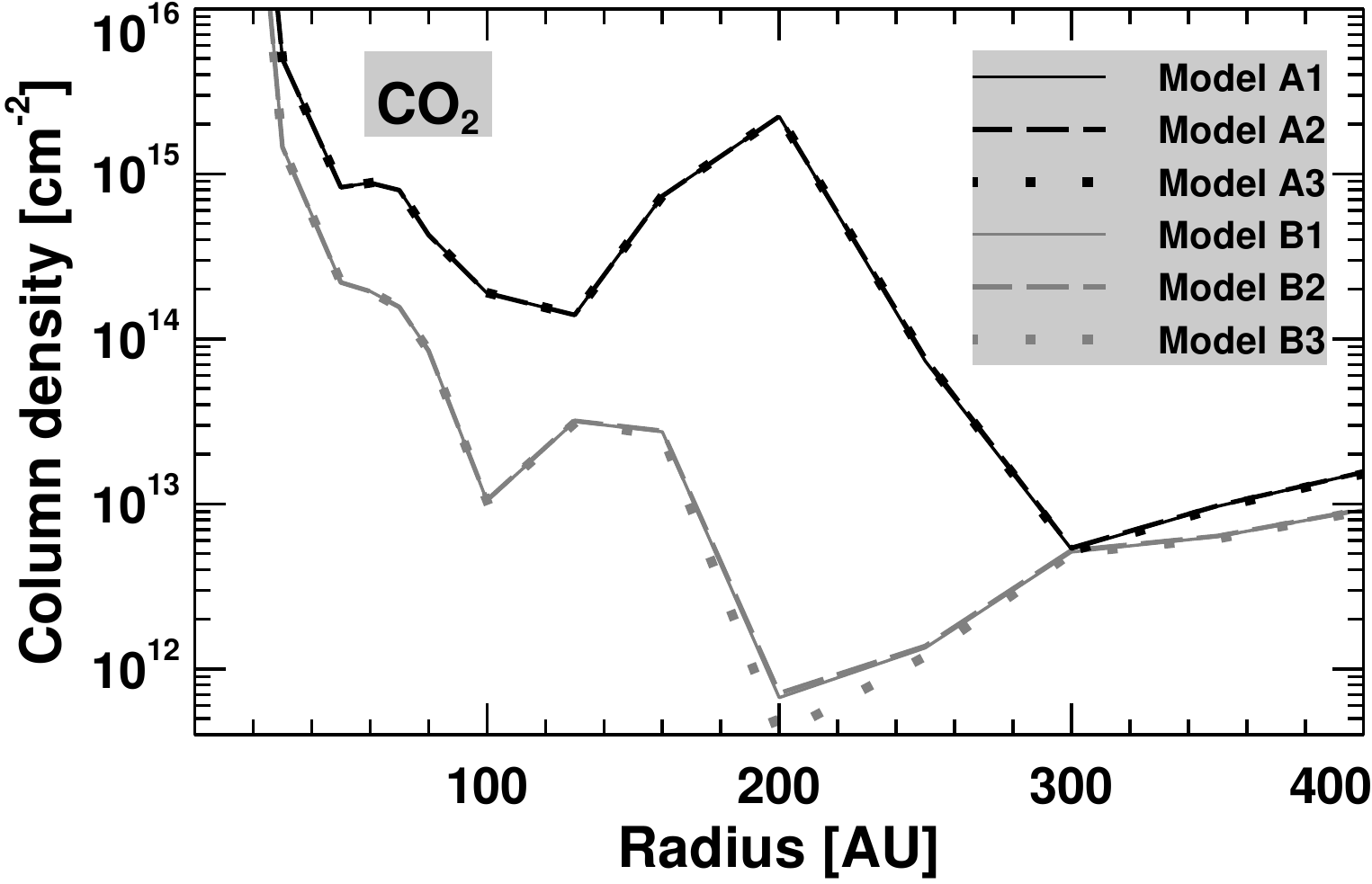}
\caption{Same as Fig.\ref{allmols_nosed} for CO$_2$ molecule.  \label{CO2_nosed}}
\end{figure}

Figure~\ref{allmols_nosed} shows the predicted vertical column densities for these species as a function of radius to the central star  for models A(1 to 3) and B(1 to 3). Here the molecules are assumed to be optically thin so that the theoretical vertical column densities are multiplies by a factor of two to represent the two sides of the disk (on both sides of the mid plane).
For most molecules, all model results with the 2-phase mode overlap. The same is found for the model results with the 3-phase mode. This result shows that the species column densities, in the ranges of radius considered here, do not depend on the initial conditions in that model. This result is not true for the main ice constituents that present significant differences among 2-phase and 3-phase models for the inner 100 au. The 2- and 3-phase models give significant differences for most molecules except for CN, CS, and H$_2$S. \\
The difference in the CO gas-phase column density comes from the fact that in the 2-phase model, CO is produced in the molecular layer (between $z/h$ = 2 and 3) by the photodissociation of gas-phase CO$_2$, which is much more abundant with the 2-phase model. Fig.~\ref{CO2_nosed} shows the computed radial column density of CO$_2$. Between 100 and 200 au in the molecular layer of the disk, we found that the origin of gas-phase CO$_2$ is either chemical desorption during its formation on the surfaces through 
the reaction $s-$CO + $s-$O or photodesorption of surface CO$_2$ (either formed by $s-$CO + $s-$O or by $s-$HCO + $s-$O ; $s-$ means the species physisorbed at the surface of the grains). Both the two desorption mechanisms are more efficient in the 2-phase model, where all surface molecules are allowed to desorb equally, explaining the much higher abundance of CO$_2$ in this model. Going outward from the star, the gas and dust temperature decreases as well as the UV flux. As a consequence, the CO$_2$ production on the grains and further desorption decreases. At 300 au, CO$_2$ is then mostly produced by the gas-phase reaction O + HCO $\rightarrow$ CO$_2$ + H in both 2- and 3-phase models. \\
The fact that the CO$_2$ abundance is much smaller in the 3-phase model, as compared to the 2-phase, at 200 au also explains the smaller SO abundance there. Indeed, SO is mostly formed by the reaction O + HS $\rightarrow$ H + SO and atomic oxygen is obtained by the photodissociation of CO$_2$.\\
The dramatically lower N$_2$H$^+$ column density with the 3-phase model is due to the fact that N$_2$ (precursor of N$_2$H$^+$) cannot evaporate from the grains while it is photodesorbed in the 2-phase model. At 200 au, the larger column density of HCN with the 2-phase model is due to the fact that HCN is formed efficiently on the surfaces and is desorbed by UV photons. \\
 The larger CCH column density in the 3-phase model, between 100 and 300 au, is due to a larger abundance peak of this molecule at $z/h$ = 2 (while the molecular abundance peaks only at $z/h$ = 3 in the 2-phase model). This difference is due to a larger abundance of s-CH$_4$ (because less carbon is locked in CO and CO$_2$) in the 3-phase model at this altitude and these radii. The larger s-CH$_4$ abundance produces the following paths: $\rm s-CH_4 + s-CCH \rightarrow s-C_2H_2$, $\rm s-C_2H_2 + s-H \rightarrow s-C_3H_2$, $\rm s-C_3H_2 + s-H \rightarrow C_3H_4$, $\rm C_3H_4 + h\nu \rightarrow C_2H_2 + H_2$, and $\rm C_2H_2 + h\nu \rightarrow CCH$. This difference is at the origin of the higher (although moderate) CCS column density in the 3-phase model as CCS forms from CCH + S.

\subsection{With grain growth and dust settling (models C to E)}

\begin{figure*}
\includegraphics[width=0.9\linewidth]{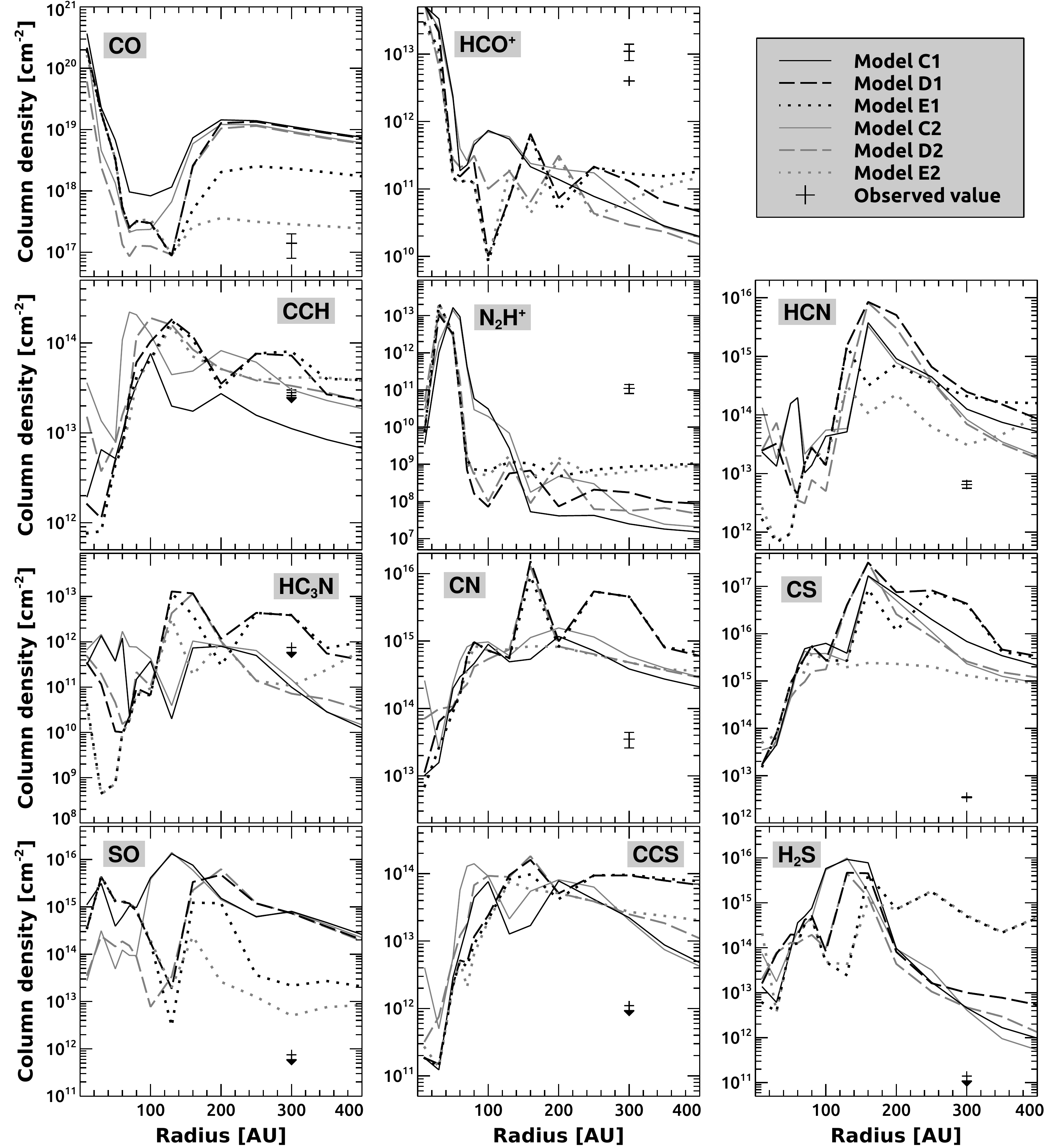}
\caption{Column densities of a selection of gas-phase species as a function of radius to the central star for models C(1 and 2), D(1 and 2) and E(1 and 2). The points at 300 au represents the observed column density. \label{allmols_sed}}
\end{figure*}

\begin{figure}
\includegraphics[width=1\linewidth]{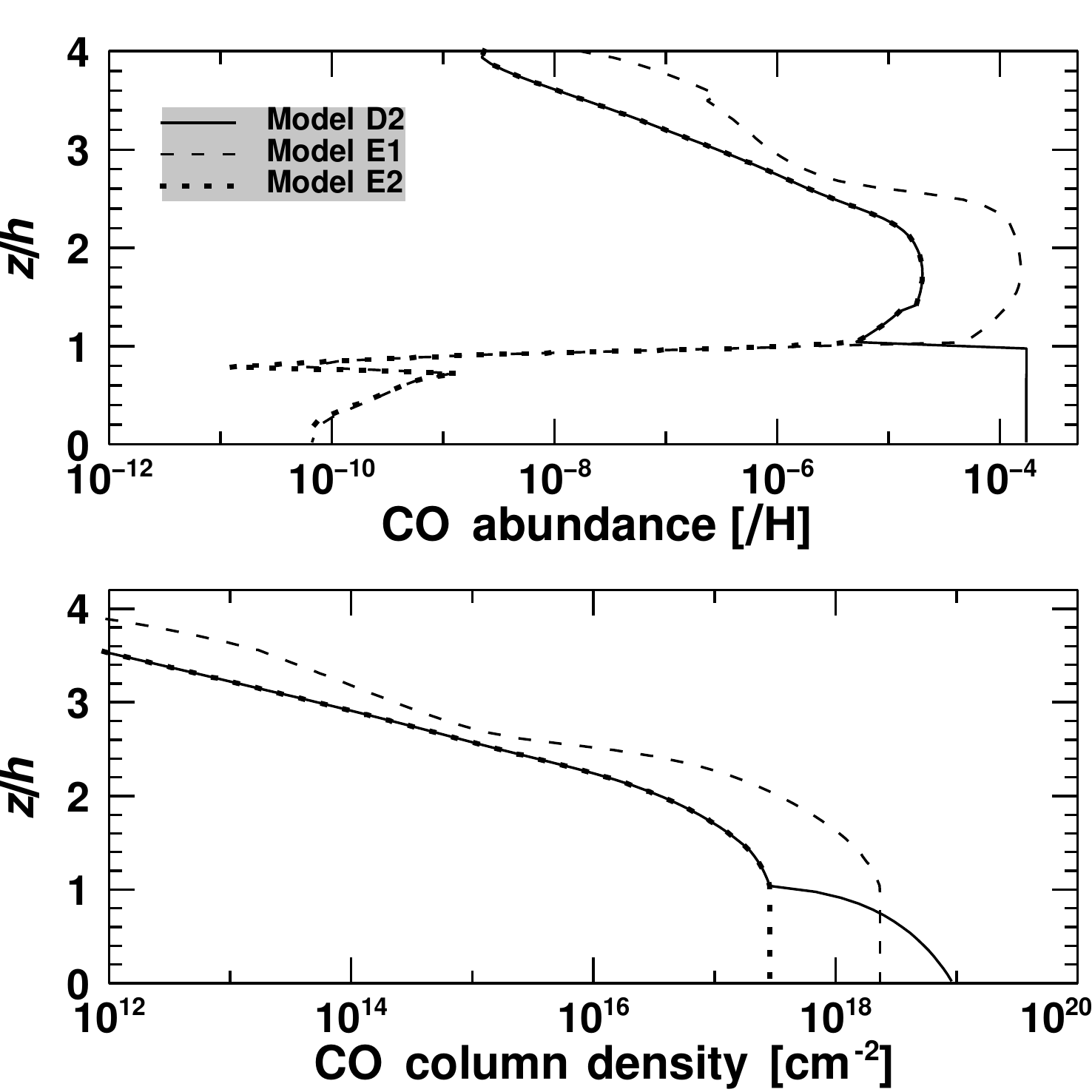}
\caption{Upper panel: CO vertical abundance at 300 au. Lower panel: vertical cumulative column density.  \label{COab}}
\end{figure}

We now assume that grains have settled (dust-to-gas mass ratio is $10^{-3}$ above $z/h$ = 1) and that the grains in the mid plane have grown (to 30$\micron$). The species column densities computed with these models are shown in Fig.~\ref{allmols_sed}. The main first result is that all species column densities depend on these assumptions. The effect depends on the species and radius but the differences can be orders of magnitude. \\
Considering CO at 100 au, the column densities spread over one order of magnitude (between $10^{17}$ and $10^{18}$~cm$^{-2}$). Beyond 200 au, the column densities are rather similar for all models and higher by two orders of magnitude than in the case of homogeneously distributed 0.1$\micron$ grains. To understand what is happening, we also run a model in which we assume dust settling but not grain growth, and two initial conditions (molecular and atomic) (models E1 and E2). In these cases, the CO column density is smaller but highly dependent on the initial conditions. When starting with molecular initial conditions (model E2), we obtain similar column densities as in the case without settling (model B3) while starting with atoms (model E1), we end up with ten times higher values (but lower than in the case with grain growth - model D1). Fig.~\ref{COab} shows the vertical CO abundance and cumulative column density computed by the two models with settling and without grain growth (models E1 an E2). Superimposed is the result of the model with settling and grain growth (and molecular initial conditions - model D2). In the models starting with molecular initial conditions, a significant fraction of the carbon is already locked as CO on grains at the beginning of the simulations. This explains the fact that the peak abundance of gas-phase CO around $z/h$ = 2 is much larger while starting from atoms. The fact that the CO column density is much smaller in the case without grain growth (similar in fact to the case with settling presented in section~\ref{without_sed}) is caused by the increase of the number of grains in the mid plane changing the depletion of CO. Indeed, the grain growth increases the cross section of individual grains hence increasing the probability of collision between gas-phase chemical species and an individual grain. Keeping the total mass constant, grain growth decreases the total number density of grains and so the probability of collision with a population of grains. The abundance (with respect to H) of 30$\micron$ grains is only $6.8\times 10^{-20}$ whereas it is $1.8\times 10^{-12}$ in the case of 0.1$\micron$ grains. 

The N$_2$H$^+$ predicted column densities peak inside 80 au and then is smaller than $10^9$~cm$^{-2}$ outward whatever the model. In the case of settling, the very low N$_2$H$^+$ column density is due to an efficient destruction by electronic recombination in the mid plane of the disk. Contrary to the case without settling, the UV penetration is more efficient and the electrons are efficiently produced by the reaction S + h$\nu$ $\rightarrow$ S$^+$ + e$^-$. This result depends on the elemental abundance of sulphur and in our case, we obtain such a strong effect because we have assumed a cosmic value. The electron fractional abundance is about $10^{-6}$ in the case of settling while it is around $10^{-10}$ in the case without settling at 300 au in the mid plane. The presence of N$_2$H$^+$ at 300 au in the DM Tau disk may then be an indication of a low electron abundance either due to a low UV penetration or a smaller sulphur abundance. 

It is quite difficult to compare the column densities computed with and without settling as the settled models produce very different results. For H$_2$S for instance, the effect depends on the radius. Inside 200 au, settling (with and without grain growth) produces larger H$_2$S gas-phase column densities. The reason for this is that the depletion of molecules on the grain surfaces in the molecular layer of the disk is smaller in the case of settling because there are less grains. At radius larger than 200 au, the settling models with grain growth produce smaller column densities because H$_2$S is produced on the grains and desorbed by exothermic surface reactions. The smaller number of grains in the grain growth models results in a smaller production of this species. The case with settling without grain growth produces column densities similar to the cases without settling. Although the number density of grains is smaller above z/h = 1, the number density below is unchanged. The region of the disk below this altitude then contributes mostly to the H$_2$S column density. 
% SO column density seems to show opposite behavior. CS should show similar trends as SO except that the column density predicted by the model with settling and no grain growth depends on the initial conditions. The molecular initial conditions produces smaller column densities, compared to the atomic ones. Similarly to CO, the smaller gas-phase column densities are due to the fact that the species are already depleted at the beginning of the disk simulations and they are trapped in the mid plane because of the small grains. HCN seems to follow a similar behavior while it seems difficult to deduce trends for CN, CCS and HC$_3$N.

\subsection{Varying the cosmic-ray ionization rate}

Only the column density of N$_2$H$^+$ is changed significantly when the cosmic-ray ionization rate is decreased (models F1 and F2 in Table~\ref{models}). Decreasing its value by two orders of magnitude decreases the N$_2$H$^+$ column density at 300 au by more than one order of magnitude. The main reason is that the H$_3^+$ column density is also decreased as N$_2$H$^+$ is formed from the reaction N$_2$ + H$_3^+$ $\rightarrow$ N$_2$H$^+$ + H$_2$. 

%\subsection{Time evolution}

%The chemistry in protoplanetary disks is often considered to be at steady state. In our models without sedimentation, the 
%BLBLA 

\section{Comparing with observations}\label{discussion}

\subsection{Observed and modeled column densities}\label{comp_obs}

\begin{table}
	\centering
	\caption{Variations of model E2 (3-phase, molecular initial conditions for a cloud age of $10^6$~yr, settling of grains without grain growth, $\zeta$ = $10^{-17}$).}
	\label{models_v2}
	\begin{tabular}{lccccc} % four columns, alignment for each
		\hline
		\hline
E2.1 & higher altitude transition $z/h$ = 2 \\
E2.2 & lower altitude transition $z/h$ = 0.5 \\
E2.3 & denser initial cloud density $2\times 10^5$~cm$^{-3}$ \\
E2.4 & S and N elemental abundances / 10 \\
E2.5 & S and N elemental abundances / 100 \\
E2.6 & E2.3 with grain growth\\
		\hline
	\end{tabular}
\end{table}

On the model predicted column densities (Figs~\ref{allmols_nosed} and \ref{allmols_sed}), we have reported the observed values at 300 au listed in Table~\ref{table_obs}. Despite the strong dispersion of model results, some of the observed column densities cannot be reproduced by any of our models at 300 au. HCO$^+$ is always underproduced while CN, HCN and all S-bearing molecules are overproduced. In the case of N$_2$H$^+$, only models without any settling can produce significant amounts of this molecule at 300 au. Using the settled model (i.e. assuming that settling is occurring in DM Tau protoplanetary disk at 300 au), we have done several other tests to see if we can come closer to the observations. Let's assume that our most realistic model is the one with settling (occurring at $z/h$ = 1) but without grain growth, starting from an initial cloud with an age of $10^6$~yr, using the 3 phase model (i.e. model E2). For the dust characteristics, one would expect grain growth to have occurred in the disk however, our models show that we need small grains in the mid plane to deplete CO, otherwise, the CO column density is much higher than the observed one. A more realistic model would require the use of a grain size distribution with both large and small grains. This is the subject of a forthcoming paper (Gavino et al. in prep).
The column densities obtained at 300 au with this model are listed in Table~\ref{table_obs} along with the observations. We can see the clear underestimation of HCO$^+$ and N$_2$H$^+$, while CN, CS, H$_2$S and CCS are overestimated by more than a factor of ten. We have run a number of additional models (starting from model E2) to test some other parameters. These models are summarized in Table~\ref{models_v2} and their results are described below.\\
Starting from model E2, we changed the altitude at which the settling occurs: E2.1 (at $z/h$ = 2) and E2.2 (at $z/h$ = 0.5). Since not all neutral abundances peak at the same altitude, the effect is not the same for all species. CS column density for instance is increased in E2.2 while H$_2$S is decreased. The two ions are increased in E2.1  but are still underestimated by two orders of magnitude. \\
One way of decreasing the species gas-phase column densities would be to start with a more depleted gas-phase composition. This can be obtained by increasing the density of the parent cloud. We tested a ten times more dense initial cloud and the resulting column densities are listed as E2.3 in Table~\ref{table_obs}. The H$_2$S and SO column densities are approximately decreased by a factor of ten whereas the ones of CS and CCS are not significantly changed. CN and HCN are not sensitive while HC$_3$N is increased, becoming higher than the observed upper limit. The HCO$^+$ column density is strongly increased (becoming close to the observed value) but the value for N$_2$H$^+$ is dramatically decreased. \\
Another way of decreasing the N- and S-bearing species column densities is to decrease the elemental abundances of N and S in the simulations. We tested depletion factors of 10 (E2.4) and 100 (E2.5). If the predicted S-bearing column densities at 300 au are almost linear to the elemental abundance of sulphur, this is not the case for CN and HCN. Decreasing the initial elemental abundances by a factor of 10 have an effect similar to the previous case (E2.3 - ten times denser initial cloud). Decreasing them by two orders of magnitude goes in the same direction and in this case, even CS and CCS are decreased. Only the ions HCO$^+$ and N$_2$H$^+$ are not changed between models E2.4 and E2.5. In models E2.3 to E2.5, the CO column densities are larger by a factor of ten as compared to the observed value. This is due to the increase in the HCO$^+$ abundance, which produces extra CO upon dissociative recombination.\\
We did one last model (E2.6) similar to E2.3 but with grain growth. In that case, CO is strongly overestimated (more than in the other cases as we have no small grains to depleted gas-phase CO), as are the ions. Only H$_2$S is getting closer to the observations as there are less grains to produce it. 

\subsection{N$_2$H$^+$ and HCO$^+$}

In our models, we have found that N$_2$H$^+$ was highly sensitive to some of the model assumptions. The use of the 3-phase model prevents the evaporation of N$_2$, precursor of N$_2$H$^+$, in the gas-phase. In the models with grain settling, this effect is  weakened as there are less grains to deplete gas-phase N$_2$. Dust settling however also increases the electron donor abundances (atomic carbon and sulphur) in the molecular layer of the disk. There are then more electrons to destroy N$_2$H$^+$. We have looked at the effect of many other parameters, such as elemental carbon and sulphur depletion (up to two orders of magnitude), initial cloud density (larger values to get more depletion of the electron donors), smaller UV irradiation field (to produce less electrons), higher or smaller altitude transitions for the dust settling, more efficient N$_2$ photodesorption. All these models produce an underestimation of the N$_2$H$^+$ column density because the electron abundance between $z/h$ = 1 to 2 is too large. The most promising models are the ones with C and S depletion, i.e. with a smaller abundance of electron donors. These models however underproduce the CO column density. 
 All our models underestimate the HCO$^+$ column density at 300 au, except for the models in which the elemental abundance of sulphur is decreased (E2.3, E2.4, and E2.5). The peak of column density for this ion is obtained at $z/h$ = 1. At this altitude, this ion is mostly produced by the reaction CO + H$_3^+$ $\rightarrow$ HCO$^+$ + H$_2$ while it is destroyed by electronic recombination. Lowering the electron donor abundance, i.e. sulphur, increases significantly the HCO$^+$ column density. \\
Both observed N$_2$H$^+$ and HCO$^+$ large column densities at 300 au are an indication of low electronic fractionation at this radius of the observed disks and at the altitude these two species peak  (i.e. between $z/h$ = 1 and 1.5). At these altitudes, the main cation is C$^+$ (it is H$_3^+$ below), orders of magnitude more abundant than N$_2$H$^+$ and HCO$^+$. The location of the C/C$^+$ vertical transition could then be key for N$_2$H$^+$ and HCO$^+$.
The very low N$_2$H$^+$ column density obtained with the 3-phase models is due to a very efficient depletion of N$_2$ from the gas, an effect much less efficient in the 2-phase model. The representation of prefectly spherical grains in these models, even for the grains that undergo growth, is very likely a crude approximation. \citet{2013A&A...557L...4K} proposed a fluffy growth mechanism for the dust in protoplanetary disks. Such fluffy grains would produce inhomogeneities in the disk dust distribution and very likely more UV penetration towards the mid plane of the disk. From a chemical point of view, fluffy grains would have a larger cross section for collisions with gas-phase species as compared to large spherical grains. In addition, the surface layer would be larger, i.e. offering more possibility for N$_2$ desorption. The models presented in this paper do not include any X-ray chemistry, which should affect the abundance of these two ions \citep{1997ApJ...480..344G,1999A&A...351..233A,2015A&A...574A.137T,2015ApJ...799..204C}.

\subsection{The CCH rings}

%\begin{figure}
%\includegraphics[width=1\linewidth]{plot_sed_CCH.pdf}
%\caption{Column densities of gas-phase CCH as a function of radius to the central star for models C(1 and 2), D(1 and 2) and E(1 and 2). \label{CCH}}
%\end{figure}

The CCH molecule has been observed by \citet{2016ApJ...831..101B} in the DM Tau protoplanetary disk. The emission of this molecule was found to form two rings: one peak around 50 au and one around 350 au. 
%Comparing our model results with these observations are difficult as no observed column density is provided in the paper. 
Some of our models with settling provide a bimodal profile with a peak between 50 and 120 au and a weaker one after 200 au (see Fig.~\ref{allmols_nosed}). The exact location of these peaks depends on the model and the maximum difference (still in column density) between the peaks values and the hole in between is less than a factor of 10. The models without settling do not give these bimodal distributions. In these cases, the CCH column density increases towards the outside until 130-160 au, remains approximately flat until 250 au and then increases again. All models, with and without settling predict a drop of CCH column density inside the 50 central au, simply due to photodissociation. It is interesting to see that the models, which produces nicer bimodal radial column densities are the ones obtained with the 3 phases model, dust settling (with or without grain growth) and with initial atomic conditions. The same models starting from molecular conditions do not give the second peak. Using a gas-phase model, \citet{2016ApJ...831..101B} argued that the CCH production can only be efficient with a gas-phase C/O ratio larger than 1. Such elemental ratio would be explained by a depletion of oxygen on the grains before the formation of the disk and driven to the mid plane of the disk. In other words, this oxygen would not participate to the chemistry in the region where they observe CCH. Such model would be close to our settled model in which we start with molecular initial conditions. In our case we have used a standard C/O ratio of 0.6 and we obtain a CCH column density larger than $10^{13}$~cm$^{-2}$ for radii larger than 100 au in all our settled models. We did not consider radial drift of the grains but as CCH abundance peaks above $z/h$ = 2, this should not affect the column density of this molecule. A more detailed work, including radiative transfer calculations would be required to conclude on this matter.

\section{Comparisons with other disks models}\label{comp_models}

We will restrain the comparison to gas-grain models for protoplanetary disks such as DM Tau and studying the effect of grain growth and/or settling. \citet{2011ApJ...726...29F} studied the importance of dust settling (without grain growth) with a 2-phase gas-grain model. Although their physical and chemical model is quite different, we can compare their computed column densities listed in their Table 2 (second lines without Ly$\alpha$ radiation, at 250 au) with our models A and C. Looking at tendencies, we obtain similar trends for CO, HCN, CN, and CCH: more column density with settling. For N$_2$H$^+$, we also find a decreased column density with settling. In our case, the decreasing HCO$^+$ column density is less obvious. We obtain column densities of the same order (for the species shown in the two papers), except for N$_2$H$^+$, which seems to be much less abundant in our settled case while the other nitrogen bearing species are more abundant. \\
Using a similar chemical model, \citet{2011ApJ...727...76V} studied the effect of both the dust settling and growth, varying in the radial and vertical directions. If their settled model (GS) produces approximately the same gas-to-dust mass ratio at 100 au as ours, their grain growth is much less important. At this radius, the maximum size increase of the dust by a factor of a few above 0.1$\micron$, i.e. much smaller than our grain growth model. In that condition, they find that the effect of dust evolution is limited at large distances from the star. Since we do not consider the radial dependency of the dust properties, we do not see this effect. The CO column density in their disk with evolved dust (i.e. with settling and grain growth) is much smaller than in all our models with settling (with or without grain growth) while the CO column density in the case with dust evolution is similar to ours in models A, except at 10 au. Secondly, N$_2$H$^+$ and CN do not appear to be sensitive to the dust evolution in their case whereas they are strongly sensitive in ours. Comparing their column densities at 100 au with ours for their A5 model (i.e. without settling and grain growth) with our model A1, we obtain similar CO, much higher N$_2$H$^+$, CN, HCN, and CCH while much lower H$_2$S and HC$_3$N.
Following this work, \citet{2013ApJ...766....8A} used a similar model but improved the treatment of the UV penetration more consistent with the evolving dust properties. Despite the fact that the values of the computed column densities are different from \citet{2011ApJ...727...76V}, they find similar trends.

\section{Conclusions}

In this paper, we have made several chemical models of the protoplanetary disk around DM Tau in order to understand the effect of several model assumptions and try to identify general trends while comparing these models to observed column densities. For this, we used a physical structure mostly based on observed parameters and assuming hydrostatic equilibrium. This 2D structure is then used as input to the Nautilus gas-grain model. We investigated the effect of the following model approximations: 2- or 3-phase model (2-phase means that all species on the grains have the same behavior whereas 3-phase means that the top 2 monolayers are chemically different from the rest of the mantle), the initial conditions (atomic, molecular with different ages), cosmic-ray ionization rate, and grain growth and dust settling. The results of a large number of models were compared with observed column densities at 300 au, published in the literature. 
Our main results are the following:
 
\begin{itemize}

\item[-] Without dust settling, the 2- and 3-phase models differ but the initial conditions have no influence. 
\item[-] With dust settling, molecular abundances become very sensitive to other model assumptions.
\item[-] Both grain growth and settling reduce the molecular depletion (and also enhances UV penetration). This boosts the CO gas-phase abundance at radii larger than 100 au. 
\item[-] Only N$_2$H$^+$ is affected by the cosmic-ray ionization rate.
\item[-] N$_2$H$^+$ is the most sensitive species. Its production requires enough N$_2$ in the gas-phase (inhibited by the 3-phase model, but boosted by the settling) and a low electron abundance (favored by low C and S elemental abundances). 
\item[-] Two key aspects of protoplanetary disk chemistry, yet to explore, are the inhomogeneities within these disks and the fluffiness of the grains.
\end{itemize}

The main conclusion of this work is that the computed chemical composition of protoplanetary disks is very sensitive to both the intrinsic chemical parameters of the models and the physical structure of the disk. In the case of the physical structure, there are still large uncertainties and the current observations are still limited leading to approximations (e.g. observed inhomogeneities, rings and spirals which are not yet properly incorporated in physical models). 
However, with future observations, utilizing the high spatial resolution of ALMA, theses uncertainties will decrease. On the chemical model itself, there are uncertainties in the chemical data but also on some of the processes \citep[not necessarily studied here such as the chemical desorption,][]{2017MolAs...6...22W}. As a consequence, it seems very difficult to accurately derive chemical parameters from observed surface densities in disks, without being very cautious.  Protoplanetary disks intrinsically exhibit complex 3D structures (observed large density and temperature gradients, small scale/unresolved dust heterogeneties...) and therefore, they are not the best objects where chemical processes can be checked and studied. Studying poorly known or new chemical processes in simpler objects with 1D structure, before incorporating them in disk chemical models, is more efficient and may provide more quantitative results.

\section*{Acknowledgements}

VW research is funded by an ERC Starting Grant (3DICE, grant agreement 336474). This study has received financial support from the French State in the frame of the "Investments for the future" Programme IdEx Bordeaux, reference ANR-10-IDEX-03-02 and the CNRS program Physique et Chimie du Milieu Interstellaire (PCMI) co-funded by the Centre National d'Etudes Spatiales (CNES).  LM acknowledges support from the NASA postdoctoral program. A portion of this research was carried out at the Jet Propulsion Laboratory, California Institute of Technology, under a contract with the National Aeronautics and Space Administration.



%\bibliographystyle{mnras}


\bibliographystyle{mnras}
\bibliography{bib}

%%%%%%%%%%%%%%%%%%%%%%%%%%%%%%%%%%%%%%%%%%%%%%%%%%

%%%%%%%%%%%%%%%%% APPENDICES %%%%%%%%%%%%%%%%%%%%%

%\appendix



%%%%%%%%%%%%%%%%%%%%%%%%%%%%%%%%%%%%%%%%%%%%%%%%%%

% Don't change these lines
\bsp	% typesetting comment
\label{lastpage}
\end{document}